# Comparing artificial frustrated magnets: tuning symmetry in nanomagnet arrays

J. Li, X. Ke, S. Zhang, D. Garand, C. Nisoli, P. Lammert, V. H. Crespi, and P. Schiffer

*Department of Physics and Materials Research Institute, Pennsylvania State University,*

*University Park, PA 16802, USA*

We study the impact of geometry on magnetostatically frustrated single-domain nanomagnet arrays. We examine square and hexagonal lattice arrays, as well as a brickwork geometry that combines the anisotropy of the square lattice and the topology of the hexagonal lattice. We find that the more highly frustrated hexagonal lattice allows for the most thorough minimization of the magnetostatic energy, and that the pair-wise correlations between moments differ qualitatively between hexagonal and brickwork lattices, although they share the same lattice topology. The results indicate that the symmetry of local interaction is more important than overall lattice topology in the accommodation of frustrated interactions.

# I. INTRODUCTION

Geometrical frustration of the interactions between atomic moments can lead to a wide range of intriguing low-temperature collective spin states, such as spin liquids, spin glasses and spin ice.[1-2] Such behavior is driven by the structure of the geometrically frustrated magnetic sub-lattice in the materials, resulting in competition between spin-spin interactions. A common thread among materials displaying exotic spin states is frustration-inducing symmetry, but a direct examination of the effect of lattice symmetry is very difficult: different lattice geometries inevitably result from chemical differences between different materials that also have important implications for the spin-spin interactions.

Artificial frustrated magnets, consisting of lithographically defined two-dimensional ferromagnetic nanostructures with single-domain elements, provide an alternative means to study geometrically frustrated magnetism. Our group and others[3-8] have examined square arrays of nanometer-scale ferromagnetic islands with perpendicular nearest neighbors in which the moment orientation resembles the 2-in/2-out spin ice state of pyrochlore materials. Additional studies[9-12] have focused upon a hexagonal geometry (equivalent to the well-known kagome lattice[13]), where the local quasi-ice vertex rule (1-in/2-out or 2-in/1-out) is strictly followed in arrays of nanowire links.[9] Since the geometries of artificial frustrated magnets are determined lithographically, lattice symmetry and topology can be directly controlled. This allows experimental investigation of a vast set of celebrated theoretical models of statistical physics such as the square-lattice Ising model and ice-type six-vertex models.[14] It has been shown that, although the moment configuration is athermal, artificial spin ice can be described through an effective thermodynamics formalism,[5] thus partially reproducing the statistical mechanics of well-known vertex models. More recent work from our group demonstrated that the moment correlations and thus the effective temperature of square artificial spin ice can be controlled by an external drive[8] that allows the system to access a wide class of microstates. Taken together, these previous studies have demonstrated that artificial frustrated magnets, while athermal, can still provide insights into a broad range of microscopic and mesoscopic statistical systems.

In the present work, we compare three lattices with independent control of lattice topology and local symmetry, while avoiding a specific framework or theoretical model. We examine the pair-wise correlations between the island moments and the energetics of the island configurations. We find that the symmetry of the local interactions is a driving force behind the accommodation of frustration, irrespective of the topology of the inter-island connectivity.

## II. SAMPLE PROPERTIES AND DEMAGNETIZATION PROTOCOL

We studied lithographically fabricated frustrated arrays of ferromagnetic permalloy islands (220 nm × 80 nm lateral and 25 nm thick), following fabrication procedures published previously.[3] The island magnetic moments are constrained to point along their long axes due to strong shape anisotropy, mimicking Ising-like spins. The coercivity of islands with these dimensions is ~ 770 Oe, independent of lattice spacing.[4] This allowed us to probe the arrays in the limits of both strong and weak interactions, i.e., small and large lattice spacings. Each array contained between 33,750 and 80,000 islands, depending on lattice spacing.

We chose the square, brickwork, and hexagonal lattice geometries for our arrays (figure 1) because they provide a simple variation in topology, i.e., inter-island connectivity. The square lattice has four-fold symmetry, with islands interacting locally at four-fold vertices. The hexagonal lattice has three-fold symmetry, with islands interacting at three-fold vertices. The brickwork lattice geometry is created by a symmetry-breaking uniaxial deformation of the hexagonal lattice to which it is topologically equivalent. Alternatively, one can form the brickwork lattice from the square lattice by removing every other island in horizontal rows. Nearest neighbors in the brickwork lattice are thus parallel or perpendicular as in the square lattice, but with the topology of the hexagonal lattice. In each of the three geometries, the magnetostatic interactions between neighboring islands at a vertex are frustrated, i.e., not all the pair-wise magnetostatic interactions can be simultaneously satisfied. Considering only the island-island interactions at each vertex (i.e., a vertex model), the hexagonal lattice has a macroscopically degenerate ground state, and the moment arrangement in figure 1 is a special case for which the vertex type is ordered as alternating 2-in/1-out and 2-out/1-in on distinct sublattices, thus lowering the further neighbor interaction energies and providing a possible ground state for this geometry.[15] By contrast, the square lattice has a two-fold degenerate ground state with no net magnetization[14] and the brickwork lattice has a two-fold degenerate ground state with a net magnetization, as depicted in Figure 1.

Because the magnetostatic energy scales in these systems are much higher than thermal energies,[3] we probed the consequences of frustration by examining the collective state of the island moments after a process of ac demagnetization. We followed our previously developed protocol for the ac demagnetization, i.e., rotating the samples in-plane while they were subjected to a stepwise deceasing in-plane external field with field polarity within the laboratory frame reversed at each step.[3-4, 8]  Initially, the external field is large enough to coerce all island moments into tracking the field. As the magnitude of the external field decreases, the island moments successively decouple from the external field, as governed by their local magnetostatics. Since the external field is decreasing in magnitude and a given island is most likely to decouple from the external field when interactions with nearby islands reinforce its current magnetization direction, it is likely that a given island moment remains static thereafter. The specified rotational demagnetization protocol generates a well-defined statistical exploration of spin configuration space wherein each island moment makes a distinct decision on its configuration relative to nearby islands, but it is not ergodic in the traditional sense. For all of the data shown below, we used the smallest accessible step size of 1.6 Oe, although data with step sizes up to 16 Oe showed qualitatively similar behavior.  After demagnetization, the island magnetic moments were imaged via magnetic force microscopy (MFM) at several locations far from the edge of each array, imaging typically 500 islands per image.  Fig. 1d, e and f show MFM images of the three different lattice geometries, with clear white and black contrast representing the island magnetic poles. Such images confirm the single-domain nature of the islands and enable us to resolve the individual magnetic orientations of the islands.  More than 3,000 islands are imaged for each data point in Figs. 2, 3, and 4, where the uncertainty in derived quantities is calculated as the standard deviation among at least five images. For all three geometries the array moment after demagnetization was zero to within experimental uncertainty.[8]

### III. RESULTS AND DISCUSSION

We first examine the magnetostatic energy of the demagnetized arrays, as quantified by summing the calculated pair-wise magnetostatic energies[16] associated with the measured moment orientations up to the seventh nearest neighbor.[8] A careful convergence study shows that truncating the summation at the seventh neighbor should introduce an error of less than 1% for macroscopically demagnetized states. Fig. 2 shows this magnetostatic energy as a function of lattice spacing for all three geometries, normalized in each case to the energy of the respective low energy states shown in Figure 1. Since the energies are negative, the normalization is to -1. All three curves of magnetostatic energy demonstrate the same monotonic decrease with decreasing lattice spacing, corresponding to the increasing influence of the magnetostatic interactions. In the normalized units, the energy of the hexagonal lattice appears to saturate at approximately -0.90, while the square and the brickwork lattices appear to approach -0.73 and -0.80 respectively. This saturation in the limit of small inter-island separation suggests that island-island interactions in this regime dominate over the other energy scales in the system. Importantly, these asymptotic energies are still well above the theoretical minimum in the limit of small lattice spacing, and the dependence upon demagnetization step size does not extrapolate to the ideal ground state energy in the limit of vanishing step size, indicating that the ground state is inaccessible, particularly for the brickwork and square lattices.[8] The differences between the lowest attainable energies for the three geometries are well outside the range of uncertainty of the data, indicating a fundamental physical difference between the lattices, i.e. the greater difficulty of kinetically accessing a ground state of lower degeneracy.

To understand this difference in energies, we next examine the correlations between neighbor moment pairs. For the purposes of analyzing the correlations, neighboring pairs are labeled in the order of their magnetostatic pair energy as S1 (square and brickwork) and H1 (hexagon) for the nearest neighbors, S2 and H2 for the next nearest neighbors, and out to the seventh nearest neighbors, as illustrated in Fig. 1(a), (b) and (c) (S4, S5, S6 and S7 pairs of brickwork lattice split into 2 subgroups respectively, $a$ and $b$ due to broken four-fold symmetry). We define a correlation value for each of the pairs as +1 (or -1) when the pair minimizes (maximizes) the magnetostatic interactions of the pair, and then average the correlation values over an entire MFM image for each geometrically distinct pair type. Fig. 3 shows the experimental values of the correlations for the first few neighbor pairs as a function of lattice spacing. As for the variation of magnetostatic energy with lattice constant, the near-neighbor correlations for the square and brickwork lattices are surprisingly similar to each other, although the lattices differ fundamentally in lattice connectivity. The large correlations for S1 and H1 at small lattice spacing is a clear reflection of the dominance of nearest neighbor interactions, consistent with previous measurements.[8] Since the brickwork lattice is a symmetry-broken variant of the hexagonal system, in the interaction-dominated saturated state at low lattice constant, the linear combination of correlations (2S1+S2)/3 for the brickwork lattice closely matches the value of H1 ~ 1/3. This congruence reflects the fact that both systems suppress the formation of maximally divergent vertices, wherein the three constituent islands point all inward or outward: i.e. they both obey the 1-in/2-out or 2-in/1-out two-dimensional ice-rule manifold. The second nearest neighbor pairs, S2 and H2 are much smaller than H1, S1, and even S3, and the S2 correlation is actually slightly negative for the square lattice at small lattice spacing. This difference is not due to weaker interactions, but instead reflects the frustration in this system, since the direct pair-wise interaction for the second neighbors is incompatible with the nearest neighbor S1 pair interactions. The correlations for the S3 neighbors are positive and much larger than the S2 correlations for both the brickwork and the square lattice, as was previous observed for the square lattice[3] and attributed to the compatibility of the S1 and S3 neighbor interactions.

The most striking difference between the geometries is in the nature of correlations between successively further neighbor pairs. Since there is no one-to-one correspondence between the different neighbor types in the different lattices, in Fig. 4 we plot the correlations for neighbor pairs up to seventh nearest neighbor as a function of the magnetostatic energy of the pair, e.g., the strength of the interaction between the island moments. As seen in the figure, for both square and brickwork arrays, the data show a sawtooth behavior, with the strength of the correlation oscillating as a function of the interaction energy between strongly positive and near-zero values. By contrast, the correlation values of hexagonal arrays show a monotonic decrease with decreasing pair energy. The qualitative difference between the hexagonal lattice and the other two geometries is consistently observed for each of the lattice spacings tested.

We attribute the difference in the decay of correlation with pair energy to the fundamental response to frustration in the different lattice geometries. Since the brickwork lattice breaks the symmetry between the three pair-wise nearest neighbor interactions of the hexagonal parent lattice, it can support more highly structured pair-wise island-island correlations. The hexagonal lattice, with its higher symmetry, can access a state in which the nearest-neighbor interactions are satisfied to the greatest extent possible *and* the further neighbor correlations decay smoothly with distance. The lower complexity of spin configuration space in this high-symmetry system facilitates this accommodation of frustration, similar to the smooth variations of spatial correlations in a spin liquid.[17] In contrast, the brickwork lattice, which has the same topology as the hexagonal lattice but breaks the point group symmetry, has a more structured distance dependence to the pair-wise correlations, with many highly unfavorable pairings. This more complex spin configuration space raises additional kinetic impediments against the action of the rotational demagnetization, sustaining a saturated, jammed final state even in the absence of macroscopic ground-state degeneracy. The more clearly structured spin-spin correlation function of the brickwork and square lattice moments is more closely analogous to the pyrochlore spin ice materials.[18] A further point of reference is provided by the highly anisotropic triangular lattice, in which the lower level of symmetry apparently leads to locally ordered domains, analogous to antiferromagnetically ordered materials.[19] Somewhat ironically, although the hexagonal system is more frustrated in the sense of having a more highly degenerate state with a vertex model, it is the most successful in approaching the ideal ground state magnetostatic energy. Apparently, the more highly degenerate ground state provides a larger target for the rotational demagnetization.

## IV. CONCLUSION AND FUTURE STUDIES

Our results demonstrate that the more frustrated hexagonal lattice is the most successful in approaching the ideal ground state magnetostatic energy. This finding leads to an important conclusion: the local symmetry of interactions in a frustrated magnet is more important than the topology of the interacting moments in determining how the system accommodates frustration. The tuning of symmetry in our experiments realizes one of the early promises of the artificial frustrated systems, in that we can perform a direct comparison between different lattices to probe how geometry impacts the resulting physics. The insight into the role of symmetry is accessible only due to the designability of artificial frustrated magnets combined with our ability to locally probe individual moments – both of these qualities are inaccessible in atomic-scale frustrated magnets. Future studies along these lines could include a more continuous variation of lattice types (e.g., a series of samples in which the angles in a hexagonal lattice are changed gradually to approach the brickwork lattice). A great deal of insight about the process of accommodating frustration could also be gained through Lorentz microscopy of artificial frustrated magnets[9] as the applied magnetic field is changed, or through time-resolved studies of colloidal systems[20] or optical trap systems[21] in which the dynamics can be probed more directly.

**ACKNOWLEDGEMENT**

We acknowledge financial support from the Army Research Office and the National Science Foundation MRSEC program (DMR-0820404) and the National Nanotechnology Infrastructure Network. We are grateful to Prof. Chris Leighton for the film deposition.

**FIGURE CAPTIONS**

FIG. 1. (Color online) The three geometries under study. (a-c) Island arrays of square, brickwork and hexagonal geometries. The white arrows show one of the ground state configurations with neighboring pairs annotated (S$n$ for square and brickwork, H$n$ for hexagon). Double arrows under each of the arrays indicate the defined lattice spacing. Because of the broken four-fold symmetry, S4, S5, S6 and S7 pairs of brickwork lattice split into 2 inequivalent subgroups respectively denoted as $a$ and $b$. (d-f) MFM images of the square array (400 nm), brickwork array (400 nm) and hexagonal array (370 nm).

FIG. 2. (Color online) Normalized array energy as a function of lattice spacing for square, brickwork and hexagonal geometries. The dashed line corresponds to the low energy states shown in fig 1 (a-c); array energies of three geometries are normalized respectively to these low energy states.

FIG. 3. (Color online) Lattice spacing dependence of correlation value between neighboring pairs referenced to their energy-minimized alignment. (a) The correlation of square (open square) and brickwork (open triangle) for S1, S2 and S3 neighbors; (b) The correlation of hexagonal (open diamond) for H1 and H2.

FIG. 4. (Color online) Correlation value of neighboring pairs as a function of dipolar pair energy obtained from micromagnetic simulation for (a) small spacing (square and brickwork geometries of 400 nm and hexagonal of 300 nm), and (b) larger spacing (square and brickwork geometries of 680 nm and hexagonal of 491 nm). The S4, S5, S6

and S7 pairs of brickwork geometry consist of *a* and *b* subgroups respectively, where subgroup *a* of S5 and S6 (*b* of S4 and S7) shows a positive correlation and subgroup *b* of S5 and S6 (*a* of S7) shows negative correlation with S4*a* near zero.

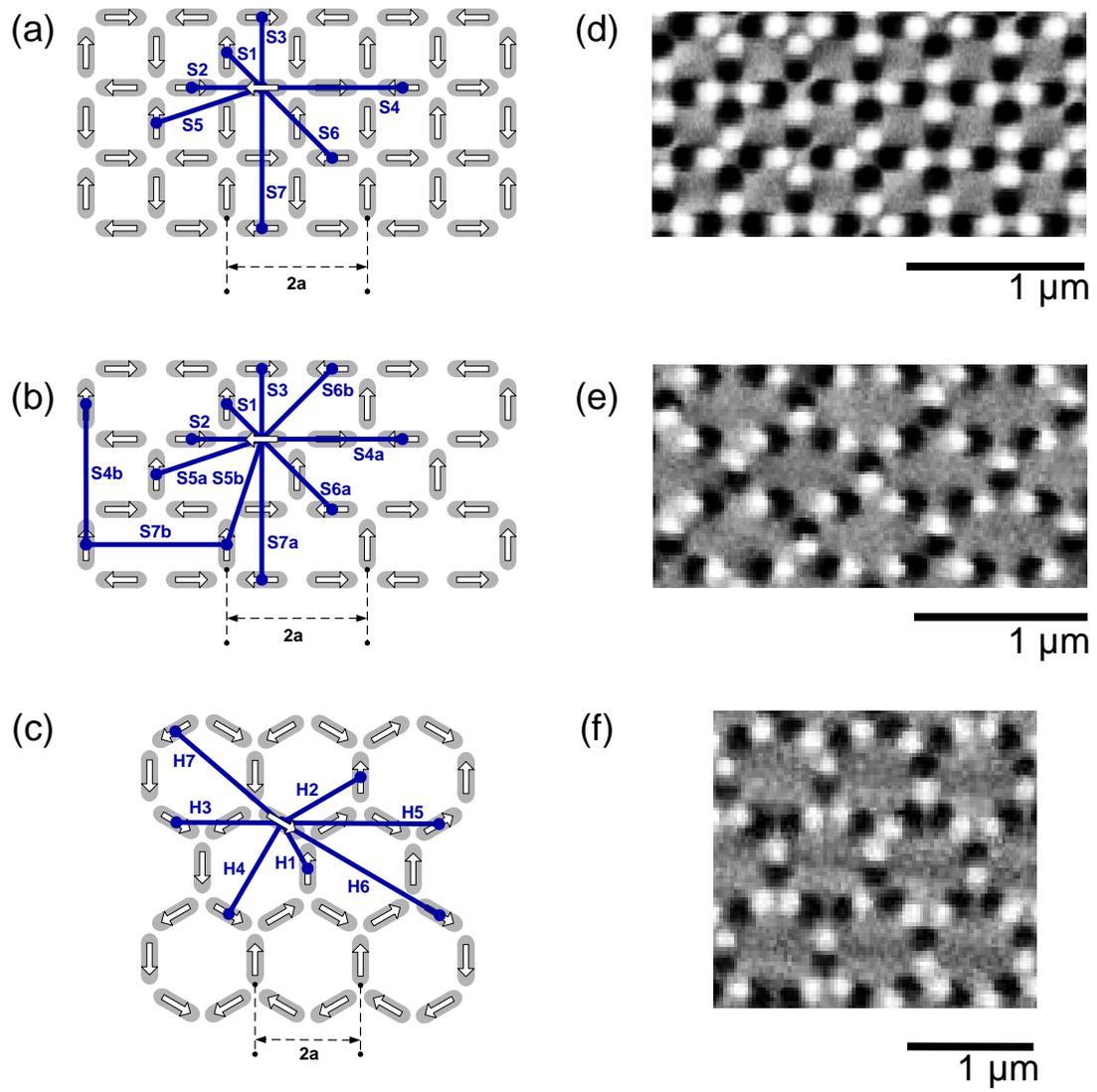

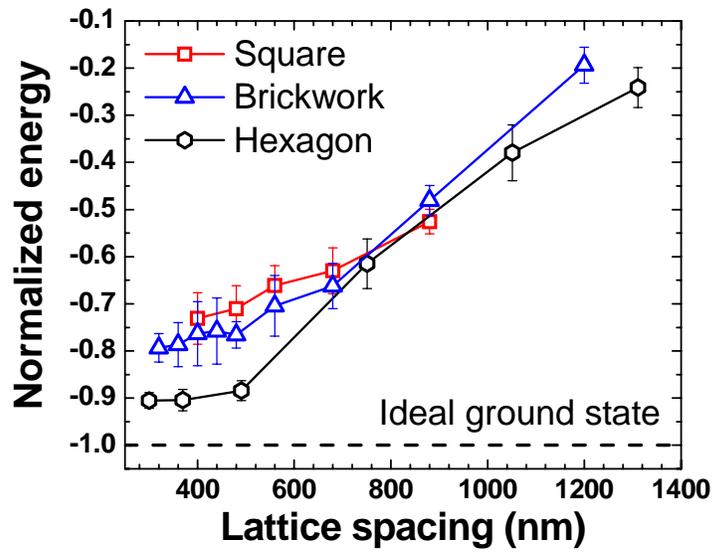

Figure 3
J. Li, *et al.*

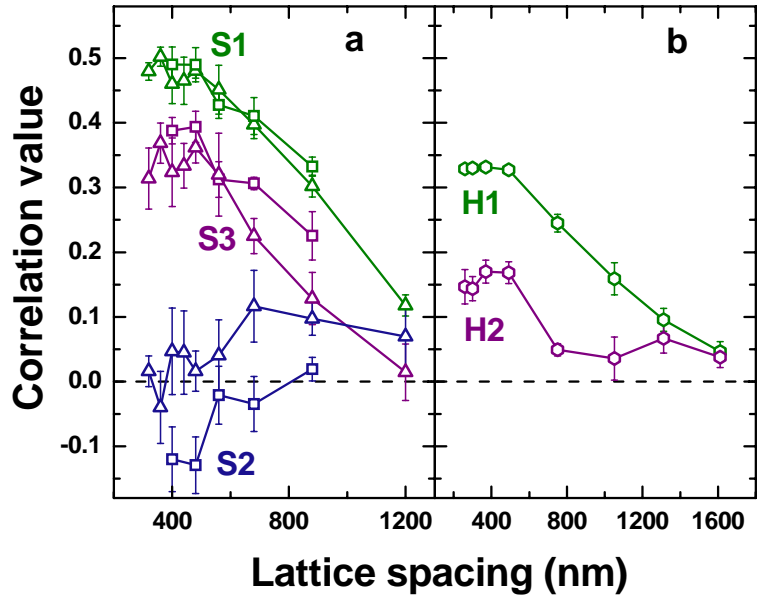

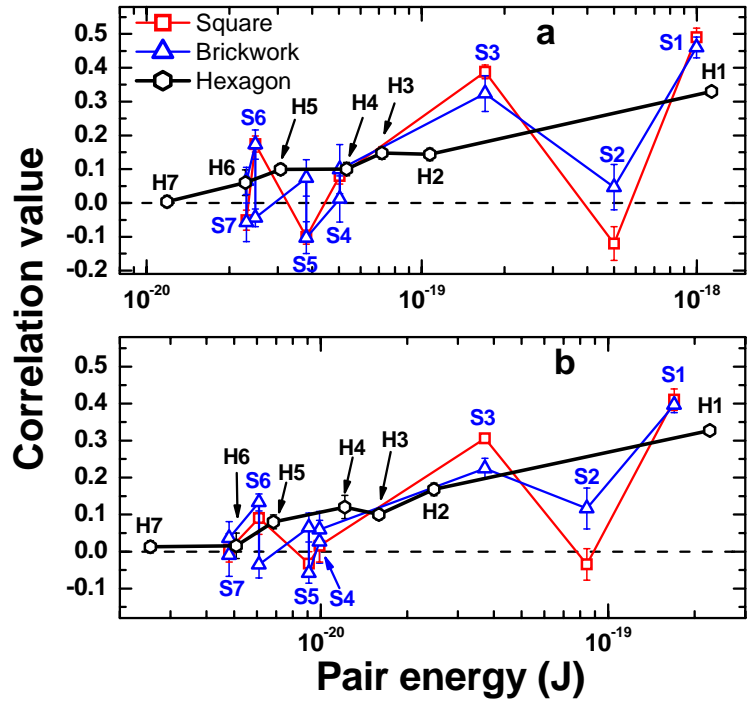